\documentclass[fleqn,12pt,twoside]{article}
\usepackage[headings]{espcrc1}

\readRCS
$Id: espcrc1.tex,v 1.2 2004/02/24 11:22:11 spepping Exp $
\usepackage{graphicx}

\newcommand{\mev}{\,{\rm MeV}}
\newcommand{\s}{\sigma}

\title{Recent results on nucleon sigma terms in lattice QCD}

\author{R.~D.~Young\address[ANL]{Physics Division, 
             Argonne National Laboratory, Argonne, Illinois 60439, USA} and
        A.~W.~Thomas\address[CSSM]{CSSM, School of Chemistry and Physics, 
             University of Adelaide, Adelaide, South Australia 5005, Australia}}

\begin{document}

\maketitle

\begin{abstract}
  It has proven a significant challenge to experiment and phenomenology
  to extract precise values of the nucleon sigma terms. This
  difficulty opens the window for lattice QCD simulations to lead the
  field in resolving this aspect of nucleon structure. Here we report
  on recent advances in the extraction of nucleon sigma terms in
  lattice QCD. In particular, the strangeness component is now being
  resolved to a precision that far surpasses best phenomenological
  estimates.
\end{abstract}

\section{NUCLEON SIGMA TERMS}
\label{sec:intro}
The nucleon sigma terms are important quantities in resolving the
dynamics of QCD, where they help to understand the role of explicit
chiral symmetry breaking in the mass of the nucleon. Indeed, the
strangeness component plays a unique role in that is purely a vacuum
polarization effect, analogous to the Lamb shift in QED. Further, the
strange quark is light enough that it probes the nonperturbative
distance scales of QCD. Beyond the fundamental importance to nucleon
structure, the strange quark condensate in the nucleon is of
significant interest in studies of the QCD phase structure at large
baryon density \cite{Nelson:1987dg} and in constraining predicted
cross sections for dark matter detection \cite{Bottino,Ellis:2008hf}.

The sigma terms of the nucleon are defined by the scalar form factors
in the limit of vanishing momentum transfer, $\sigma_q = m_q \langle N
| \bar{q} q | N \rangle$, where $q$ denotes the quark flavour of
interest. The light-quark sigma term (or pion-nucleon sigma term),
$\sigma_\ell=m_\ell\langle \bar{u}u+\bar{d}d\rangle$ (with
$m_\ell\equiv (m_u+m_d)/2$), is related to $\pi$--$N$ scattering
through a chiral low-energy relation
\cite{Cheng:1970mx,Brown:1971pn,Gasser:1990ce},
\begin{equation}
\Sigma_{\pi N} \equiv \sigma_\ell 
            = \Sigma_{\pi N}^{\rm CD} - \Delta_R - \Delta_\sigma \,,
\end{equation}
where the experimental input required is the Born-subtracted,
isoscalar $\pi N$ scattering amplitude evaluated at the (unphysical)
Cheng-Dashen point, $\Sigma_{\pi N}^{\rm CD}$.  The smallness of the
up and down quark masses ensures that both the remainder term,
$\Delta_R$ \cite{Brown:1971pn,Bernard:1996nu}, and the form factor
correction, $\Delta_\sigma$ \cite{Gasser:1990ap}, can be reliably
estimated.  An early analysis of experimental results
\cite{Koch:1982pu} led to the value $\sigma_\ell = 45 \pm 8\mev$
\cite{Gasser:1990ap}, while a more recent analysis suggested a higher
value, $64 \pm 7\mev$ \cite{Pavan:2001wz}.

Unlike the light quarks, the strange quark is too heavy to invoke a
low-energy relation.  Therefore $\s_s$ has generally been estimated by
studying the breaking of SU(3) within the baryon octet
\cite{Nelson:1987dg,Gasser:1980sb,Borasoy:1996bx}. Here, the observed
spectrum has been used to derive a constraint on the non-singlet
combination
\begin{equation}
\sigma_0 = m_\ell \langle N | \bar{u}u + \bar{d}d - 2\bar{s}s | N \rangle\,,
\end{equation}
where chiral effective field theory studies lead to a value
$\sigma_0=36\pm7\mev$ \cite{Borasoy:1996bx}, building on earlier
estimates by Gasser \cite{Gasser:1980sb}. The difference between $\sigma_0$
and the extracted $\sigma_\ell$ then defines the strange-quark
contribution, $\s_\ell-\s_0=2\frac{m_\ell}{m_s}\s_s$. Equivalently,
the relative strangeness component is often discussed in terms of the
parameter $y$,
\begin{equation}
\frac{\s_0}{\s_\ell} = 1 - \frac{2\langle N|\bar{s}s|N\rangle}{\langle N|\bar{u}u+\bar{d}d|N\rangle} \equiv 1-y \,.
\end{equation}
Taking the above pion-nucleon sigma term values and the estimated
$\s_0$ gives values $y \simeq 0.2\pm 0.2$ and $0.44\pm 0.13$, and
using a strange-to-light quark mass ratio of $\sim 25$ gives
strangeness sigma terms, $\s_s = 110 \pm 130$ and $350 \pm 120\mev$,
respectively.  This approach clearly leads to a result for $\s_s$
which is very sensitive to the precise value of $\s_\ell$. We also
note that even with perfect $\pi$--$N$ data to better constrain
$\s_\ell$, the $7\mev$ uncertainty in $\s_0$ alone leads to a $\sim
90\mev$ uncertainty in $\s_s$. This limitation clearly opens the way
for lattice QCD to offer significant improvement.

\section{LATTICE CALCULATIONS}
\label{sec:sigma}
\subsection{Light quarks}
\begin{figure}[t]
\begin{center}
\includegraphics[width=0.48\textwidth]{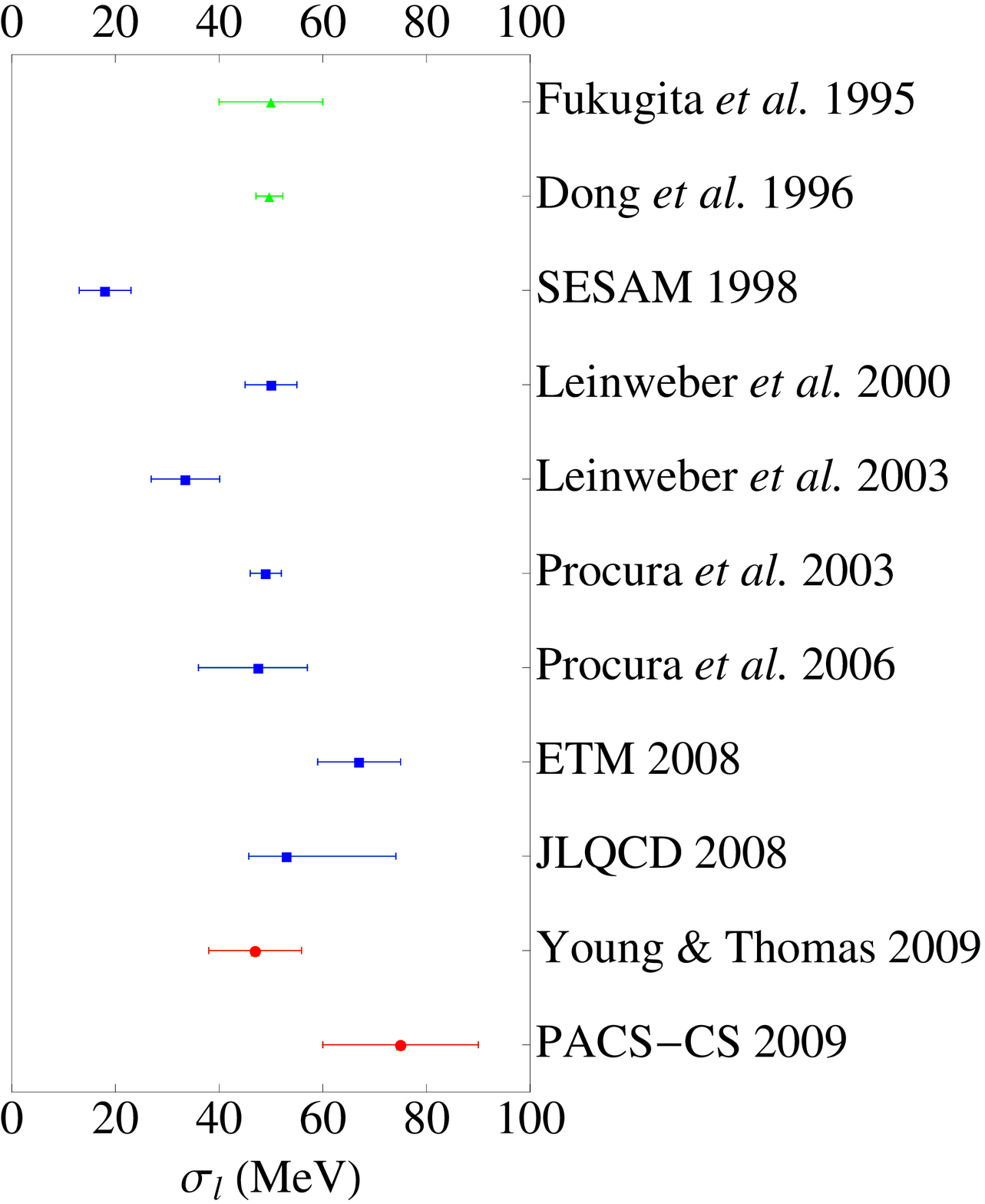}\hspace*{0.48\textwidth}
\put(-0.46,0.25){\includegraphics[width=0.48\textwidth]{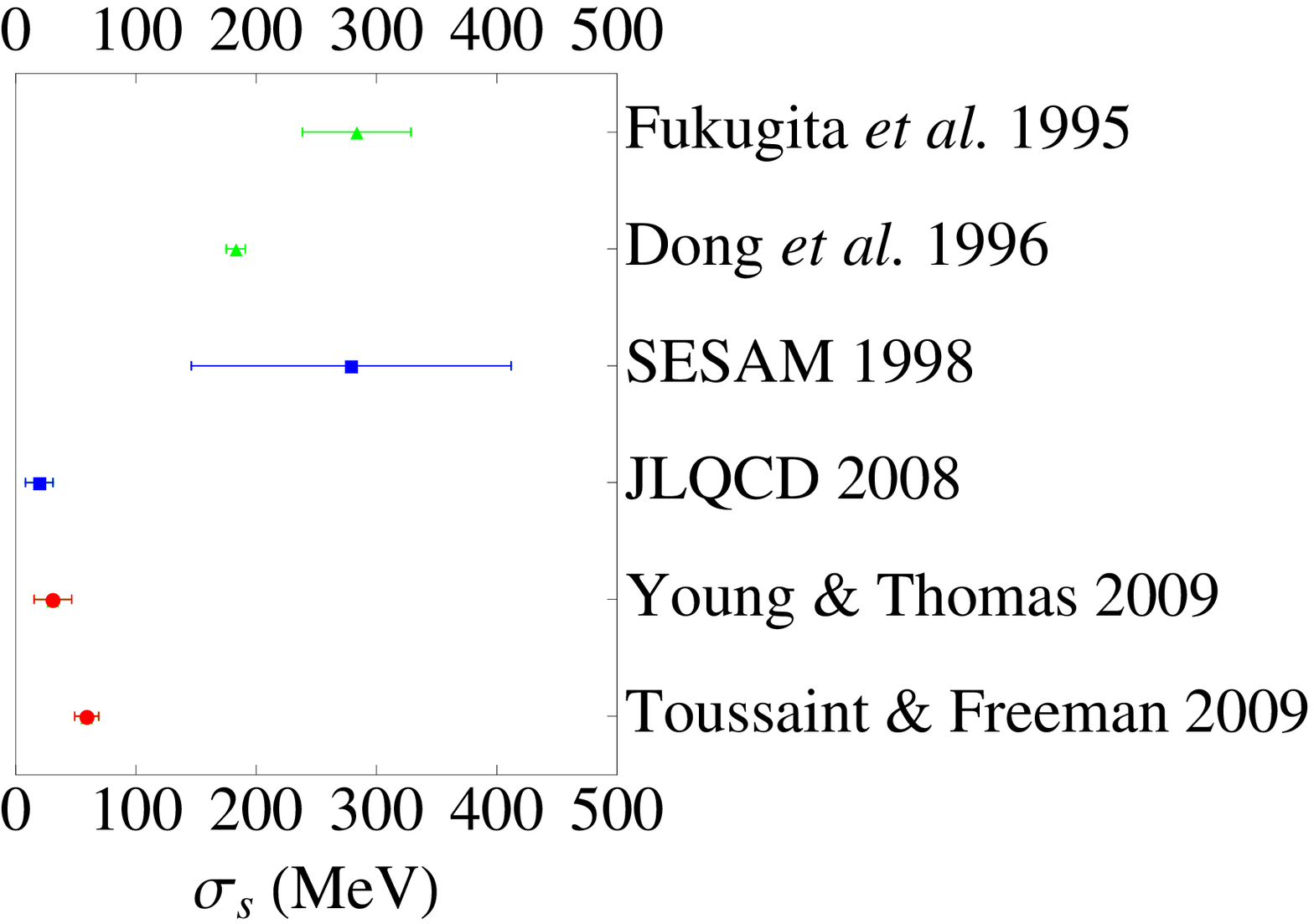}}
\caption{Summary of lattice QCD calculations of the light-quark sigma
  term (left panel) and strangeness sigma term (right
  panel). Triangles, squares and circles denote quenched, 2-flavour
  and 2+1-flavour lattices, respectively. We caution that different
  calculations have invested differing degrees of effort into
  quantifying their error estimate.
\label{fig:sig}}
\end{center}
\end{figure}

In early lattice simulations, limited computing power forced the use
of heavy quarks and the quenching of the light quarks in the theory
\cite{Maiani:1987by,Gusken:1988yi}. The first direct calculations of
the disconnected contributions (though still on a quenched gauge
ensemble), including an estimate of the strange quark contributions,
were performed in the work of Fukugita~{\it et al.}
\cite{Fukugita:1994ba} and Dong~{\it et al.} \cite{Dong:1995ec}. In
these works, the $\pi$--N sigma was found to be consistent with
observation --- see Figure~\ref{fig:sig} for a summary of results.

The first determination using (2 flavours of) dynamical quarks found a
value for $\s_{\ell}$ that is smaller than phenomenological
expectations \cite{Gusken:1998wy}. A notable feature of this
calculation is that consistent results were demonstrated between the
direct calculation of the disconnected loops and that extracted from
the fit to the nucleon mass, using the Feynman-Hellman
relation\footnote{Commonly the quark masses in this relation are
  replaced by the meson mass squared using the
  Gell-Mann--Oakes--Renner relation.}
\begin{equation}
\s_q = m_q \frac{\partial M_N}{\partial m_q}\,.
\end{equation}
While this calculation appeared to yield a value that was rather low,
it was soon identified that the sigma term is particularly sensitive
to chiral extrapolation \cite{Leinweber:2000sa}. After this time, most
extractions of $\s_\ell$ have come by application of the
Feynman-Hellman relation to the mass extrapolation
\cite{Leinweber:2003dg,Procura:2003ig,Procura:2006bj,Alexandrou:2008tn,Ohki:2008ff,Young:2009zb,Ishikawa:2009vc}.

\setlength{\unitlength}{\textwidth}

It was common in the early calculations that the quoted error only
reflected the statistical uncertainty in the underlying
extraction. The work of Procura~{\it et al.} \cite{Procura:2006bj}
made a detailed investigation of the systematic errors associated with
the input parameters of their functional form. Contrasting this with
their earlier work \cite{Procura:2003ig}, where only the statistical
error was quoted, is a useful guide as to the potential size of
systematic effects. Of course, various systematics will continue to be
better controlled as the simulation results improve.

While it is not obvious from Figure~\ref{fig:sig}(a) that the lattice
results are showing convergence --- without taking bias --- there does
appear to be consensus that the lattice results are consistent with
the two phenomenological extractions discussed above.

\subsection{Strange quark}
Early lattice estimates of the strangeness sigma term supported the
rather large values that had been inferred from phenomenology
\cite{Fukugita:1994ba,Dong:1995ec}. Going beyond the quenched
approximation, the 2-flavour dynamical calculations of the SESAM
Collaboration \cite{Gusken:1998wy} also supported the larger
strangeness values, though with substantial uncertainty. Another early
calculation highlighted some of the challenges of resolving these
matrix elements and concluded that the strangeness component is most
likely rather small \cite{Michael:2001bv}.

While the strange quark in the 2-flavour simulations is essentially
quenched, it is interesting to consider the light-quark disconnected
diagrams, which are truly dynamical. One can consider the ratio of the
disconnected to connected (or sea to valence) contributions to the
sigma term,
\begin{equation}
R_{d/c} = \frac{\langle N|\bar{u}u+\bar{d}d|N\rangle_{disc}}{\langle N|\bar{u}u+\bar{d}d|N\rangle_{con}}
\end{equation}
In the quenched simulations a relatively large value for this ratio
was observed, with $R_{d/c}=2.23(52)$ \cite{Fukugita:1994ba} and
$1.79(7)$ \cite{Dong:1995ec}. We note that the effect of unqenching
appears to reduce this ratio, with the corrsponding ratio found to be
$R_{d/c}=1.26(57)$ in the dynamical simulation
\cite{Gusken:1998wy}. This indicates a decrease in the relative
strength of the disconnected component.

Extending this discussion, a modern determination of this ratio by
JLQCD \cite{Ohki:2008ff} has found $R_{d/c}=0.41(5)$ (taking PQ-b as the
``best estimate'') --- suggesting even further reduction of this ratio
as results better describe the physical regime. This study constrained
the nucleon mass using light-quark masses taking values over the range
$m_s/6$--$m_s$ (or pion masses $\sim 290$--$750\mev$).

In a SU(3) symmetric ($m_u=m_d=m_s$) 3-flavour dynamical calculation,
the disconnected contributions of each of the three quarks are
identical, satisfying the identity
\begin{equation}
2\langle N | \bar{s}s | N\rangle_{disc} \stackrel{SU(3)}{=} \langle N | \bar{u}u + \bar{d}d | N\rangle_{disc}\,.
\end{equation}
In a 2-flavour simulation, the literal $\bar{s}s$ contribution is
zero, yet it could be reasonable to adopt the above relation to infer
the $\bar{s}s$ from a 2-flavour calculation. Indeed this is the
approach taken by \cite{Ohki:2008ff}, where at the SU(3) symmetric
point one can infer $y\sim 0.09$. Further taking the 2-flavour result
for the increase in both the connected and disconnected light-quark
terms as the light-quarks are reduced to the physical point reduces
the estimate of $y$ to $\sim 0.03$ \cite{Ohki:2008ff} (assuming that the
strangeness term is only weakly dependent on the light-quark masses).
Indeed, when combined with their extraction for the $\sigma_\ell$, this
leads to a significantly reduced value of the strangeness sigma
term $\sigma_s=20\pm 12\mev$.

The major cause of this reduction compared with the previous estimates
comes from the observation that the derivative of the nucleon mass
($\frac{\partial M_N}{\partial m_{sea}}$) is quite strongly dependent
on the sea-quark mass ($m_{sea}$). Indeed, this derivative was
observed to be enhanced by a factor of $\sim$10 in going from the
strange-quark mass to the light-quark mass. Beyond the numerical
improvements in the underlying lattice simulations, this study has
also incorporated the consequences of chiral symmetry breaking, where
there are known to be substantial chiral logarithms in these
derivatives --- leading to the observed enhancement in the light-quark
domain.

These authors have recently reported results on testing their method
against a direct calculation of the disconnected strangeness
component, finding compatible results \cite{Takeda:2009ga}. Further,
the analysis of the JLQCD Collaboration is being extended to 2+1
flavours of dynamical quarks, where preliminary results also support
their 2-flavour findings \cite{Ohki:2009mt}.

Another opportunity to study the strange-quark mass dependence came
with a series of calculations of the baryon spectrum in 2+1-flavour
dynamical simulations which emerged in 2008
\cite{WalkerLoud:2008bp,Aoki:2008sm,Lin:2008pr,Durr:2008zz} (after an
early venture in 2001 \cite{Bernard:2001av}). These new 3-flavour
results allowed tests of the SU(3) chiral expansion at the lattice
quark masses
\cite{WalkerLoud:2008bp,Young:2009zb,Ishikawa:2009vc}. Indeed the poor
convergence of the SU(3) expansion, already recognised at the physical
quark masses \cite{Borasoy:1996bx,Gasser:1980sb}, was rediscovered on
the lattice. These three chiral-lattice papers took three different
approaches to deal with this problem. PACS-CS \cite{Ishikawa:2009vc}
simply abondoned the SU(3) formulation; LHPC \cite{WalkerLoud:2008bp}
sacrificed the coefficients of the chiral logarithms in order to
stabilise the fits; we \cite{Young:2009zb} introduced a single new fit
parameter that acts to separate the low- and high-energy contributions
to the chiral loop integrals.

The results of Ref.~\cite{Young:2009zb} built on the established
benefits of using finite-range regularisation, which has previously
been shown to dramatically improve the SU(3) convergence
\cite{Gasser:1980sb,Donoghue:1998bs} and offer robust chiral extrapolation for
lattice QCD \cite{Leinweber:2003dg,Leinweber:1999ig}. The analysis of
Ref.~\cite{Young:2009zb} demonstrated the ability to extrapolate in
both the light- and strange-quark masses to accurately reproduce the
physical octet baryon spectrum, and also extrapolate to the heavier
quark masses (not used in the fits) as calculated on the lattice. The
results were found to be consistent with two different lattice
discretizations, which also allowed an estimate of potential
discretization effects.

Using the Feynman-Hellman relation, we found $\s_\ell$ to be
compatible with phenomenological and lattice estimates. And the
strangeness sigma term was identified to be relatively small,
supporting the 2-flavour calculation of JLQCD \cite{Ohki:2008ff}.

Another recent 2+1-flavour dynamical result for $\s_s$ has been
reported by Toussaint and Freeman \cite{Toussaint:2009pz}, using an
application of the Feynman-Hellman relation on the nucleon correlator.
This technique substantially differs from the approaches based on
fitting the the nucleon mass and therefore provides a strong
independent test. The result of Toussaint and Freeman
\cite{Toussaint:2009pz} is marginally higher than the values reported
by JLQCD \cite{Ohki:2008ff} and by us
\cite{Young:2009zb}. Nevertheless, as seen in Figure~\ref{fig:sig}(b),
the modern lattice results for $\s_s$ are in agreement that the size
of the strangeness sigma term is substantially smaller than has been
previously thought.

\section{IMPACT}
As described above, the dominant uncertainty in knowledge of the
scalar quark couplings has lies in the strangeness component. The
current generation of lattice QCD calculations has primarily resolved
this limitation by providing stringent new limits on the strange quark
sigma term. As a result of this improvement, the predicted cross
sections for models of dark matter can now be revised. In a recent
paper, Giedt et al.~\cite{Giedt:2009mr} have investigated the impact
of the new lattice QCD results on the cross sections for a class of
supersymmetric models. The poorly known strangeness $\s$-term had previously
resulted in cross sections which varied by roughly an order of
magnitude \cite{Ellis:2008hf}. The uncertainties have been
dramatically reduced by the lattice calculations and importantly, one
now has significant discrimination power between the ensemble of
models considered.

This work was supported by the U.S. Department of Energy, Office of
Nuclear Physics, under Contract No. DE-AC02-06CH11357; and the
Australian Research Council.

\end{document}